\documentstyle[prl,aps,preprint,epsf,floats]{revtex}

\begin{document}
\setcounter{page}{0}
\def\footnoterule{\kern-3pt \hrule width\hsize \kern3pt}
\tighten
\title{Color Superconductivity and Chiral Symmetry Restoration 
at Nonzero Baryon Density and Temperature\thanks
{This work is supported in part by funds provided by the U.S.
Department of Energy (D.O.E.) under cooperative 
research agreement \# DE-FC02-94ER40818.}}

\author{J{\"u}rgen Berges
and Krishna Rajagopal\footnote{Email addresses: 
{\tt berges@ctp.mit.edu}  and {\tt krishna@ctp.mit.edu}}}

\address{Center for Theoretical Physics \\
Laboratory for Nuclear Science \\
and Department of Physics \\
Massachusetts Institute of Technology \\
Cambridge, Massachusetts 02139 \\
{~}}

\date{MIT-CTP-2725, hep-ph/9804233}
\maketitle

\thispagestyle{empty}

\begin{abstract}

We explore the phase diagram of strongly
interacting matter 
as a function of temperature and 
baryon number density, using a class of models for two-flavor QCD
in which the interaction between quarks is modelled
by that induced by instantons.
Our treatment allows us to investigate the
possible simultaneous formation of condensates in
the conventional quark--anti-quark channel 
(breaking chiral symmetry) and in a quark--quark channel
leading to color superconductivity: the spontaneous breaking of color
symmetry via the formation of quark Cooper pairs.
At low temperatures, chiral symmetry restoration
occurs via a first order transition between a phase
with low (or zero) baryon density and a high density
color superconducting phase.  We find color superconductivity
in the high density phase for temperatures less than of order
tens to 100 MeV, and find coexisting $\langle qq \rangle$
and $\langle\bar q q\rangle$ condensates in this phase
in the presence of a current quark mass.
At high temperatures, the chiral phase transition
is second order in the chiral limit and is a smooth crossover
for nonzero current quark mass.
A tricritical point separates the first 
order transition at high densities from the second order 
transition at high temperatures. In
the presence of a current quark mass this tricritical point
becomes a second order phase transition with Ising model
exponents, suggesting that a long correlation length
may develop in heavy ion collisions in which the
phase transition is traversed at the appropriate density.

\end{abstract}

\vspace*{\fill}


\newcommand{\nwc}{\newcommand}
%
%
\nwc{\cl}  {\clubsuit}
\nwc{\di}  {\diamondsuit}
\nwc{\sps} {\spadesuit}
\nwc{\hyp} {\hyphenation}
\nwc{\be}  {\begin{equation}}
\nwc{\ee}  {\end{equation}}
\nwc{\ba}  {\begin{array}}
\nwc{\ea}  {\end{array}}
\nwc{\bdm} {\begin{displaymath}}
\nwc{\edm} {\end{displaymath}}
\nwc{\bea} {\be\ba{rcl}}
\nwc{\eea} {\ea\ee}
\nwc{\ben} {\begin{eqnarray}}
\nwc{\een} {\end{eqnarray}}
\nwc{\bda} {\bdm\ba{lcl}}
\nwc{\eda} {\ea\edm}
\nwc{\bc}  {\begin{center}}
\nwc{\ec}  {\end{center}}
\nwc{\ds}  {\displaystyle}
\nwc{\bmat}{\left(\ba}
\nwc{\emat}{\ea\right)}
\nwc{\non} {\nonumber}
\nwc{\bib} {\bibitem}
\nwc{\lra} {\longrightarrow}
\nwc{\Llra}{\Longleftrightarrow}
\nwc{\ra}  {\rightarrow}
\nwc{\Ra}  {\Rightarrow}
\nwc{\lmt} {\longmapsto}
\nwc{\pa} {\partial}
\nwc{\iy}  {\infty}
\nwc{\ovl}  {\overline}
\nwc{\hm}  {\hspace{3mm}}
\nwc{\lf}  {\left}
\nwc{\ri}  {\right}
\nwc{\lm}  {\limits}
\nwc{\lb}  {\lbrack}
\nwc{\rb}  {\rbrack}
\nwc{\ov}  {\over}
\nwc{\pr}  {\prime}
\nwc{\nnn} {\nonumber \vspace{.2cm} \\ }
\nwc{\Sc}  {{\cal S}}
\nwc{\Lc}  {{\cal L}}
\nwc{\Rc}  {{\cal R}}
\nwc{\Dc}  {{\cal D}}
\nwc{\Oc}  {{\cal O}}
\nwc{\Cc}  {{\cal C}}
\nwc{\Pc}  {{\cal P}}
\nwc{\Mc}  {{\cal M}}
\nwc{\Ec}  {{\cal E}}
\nwc{\Fc}  {{\cal F}}
\nwc{\Hc}  {{\cal H}}
\nwc{\Kc}  {{\cal K}}
\nwc{\Xc}  {{\cal X}}
\nwc{\Gc}  {{\cal G}}
\nwc{\Zc}  {{\cal Z}}
\nwc{\Nc}  {{\cal N}}
\nwc{\fca} {{\cal f}}
\nwc{\xc}  {{\cal x}}
\nwc{\Ac}  {{\cal A}}
\nwc{\Bc}  {{\cal B}}
\nwc{\Uc}  {{\cal U}}
\nwc{\Vc}  {{\cal V}}
%
%
\nwc{\Th} {\Theta}
\nwc{\th} {\theta}
\nwc{\vth} {\vartheta}
\nwc{\eps}{\epsilon}
\nwc{\si} {\sigma}
\nwc{\Gm} {\Gamma}
\nwc{\gm} {\gamma}
\nwc{\bt} {\beta}
\nwc{\La} {\Lambda}
\nwc{\la} {\lambda}
\nwc{\om} {\omega}
\nwc{\Om} {\Omega}
\nwc{\dt} {\delta}
\nwc{\Si} {\Sigma}
\nwc{\Dt} {\Delta}
\nwc{\al} {\alpha}
\nwc{\vph}{\varphi}
\nwc{\zt} {\zeta}
%
%
\def\tr{\mathop{\rm tr}}
\def\Tr{\mathop{\rm Tr}}
\def\Det{\mathop{\rm Det}}
\def\Im{\mathop{\rm Im}}
\def\Re{\mathop{\rm Re}}
\def\secder#1#2#3{{\partial^2 #1\over\partial #2 \partial #3}}
\def\bra#1{\left\langle #1\right|}
\def\ket#1{\left| #1\right\rangle}
\def\VEV#1{\left\langle #1\right\rangle}
\def\gdot#1{\rlap{$#1$}/}
\def\abs#1{\left| #1\right|}
\def\pr#1{#1^\prime}
\def\ltap{\raisebox{-.4ex}{\rlap{$\sim$}} \raisebox{.4ex}{$<$}}
\def\gtap{\raisebox{-.4ex}{\rlap{$\sim$}} \raisebox{.4ex}{$>$}}
\nwc{\Id}  {{\bf 1}}
\nwc{\diag} {{\rm diag}}
\nwc{\inv}  {{\rm inv}}
\nwc{\mod}  {{\rm mod}}
\nwc{\hal} {\frac{1}{2}}
\nwc{\tpi}  {2\pi i}
\def\contract{\makebox[1.2em][c]{
        \mbox{\rule{.6em}{.01truein}\rule{.01truein}{.6em}}}}
\def\slash#1{#1\!\!\!/\!\,\,}
%
%
\def\KK{{\rm I\kern -.2em  K}}
\def\NN{{\rm I\kern -.16em N}}
\def\RR{{\rm I\kern -.2em  R}}
\def\ZZ{Z \kern -.43em Z}
\def\QQ{{\rm \kern .25em
             \vrule height1.4ex depth-.12ex width.06em\kern-.31em Q}}
\def\CC{{\rm \kern .25em
             \vrule height1.4ex depth-.12ex width.06em\kern-.31em C}}
\def\ZZZ{Z\kern -0.31em Z}

\def\ap#1{Annals of Physics {\bf #1}}
\def\cmp#1{Comm. Math. Phys. {\bf #1}}
\def\hpa#1{Helv. Phys. Acta {\bf #1}}
\def\ijmpa#1{Int. J. Mod. Phys. {\bf A#1}}
\def\jpc#1{J. Phys. {\bf C#1}}
\def\mpla#1{Mod. Phys. Lett. {\bf A#1}}
\def\npa#1{Nucl. Phys. {\bf A#1}}
\def\npb#1{Nucl. Phys. {\bf B#1}}
\def\nc#1{Nuovo Cim. {\bf #1}}
\def\pha#1{Physica {\bf A#1}}
\def\pla#1{Phys. Lett. {\bf #1A}}
\def\plb#1{Phys. Lett. {\bf #1B}}
\def\pr#1{Phys. Rev. {\bf #1}}
\def\pra#1{Phys. Rev. {\bf A#1 }}
\def\prb#1{Phys. Rev. {\bf B#1 }}
\def\prp#1{Phys. Rep. {\bf #1}}
\def\prc#1{Phys. Rep. {\bf C#1}}
\def\prd#1{Phys. Rev. {\bf D#1 }}
\def\prle#1{Phys. Rev. Lett. {\bf #1}}
\def\ptp#1{Progr. Theor. Phys. {\bf #1}}
\def\rmp#1{Rev. Mod. Phys. {\bf #1}}
\def\rnc#1{Riv. Nuo. Cim. {\bf #1}}
\def\zpc#1{Z. Phys. {\bf C#1}}
\def\APP#1{Acta Phys.~Pol.~{\bf #1}}
\def\AP#1{Annals of Physics~{\bf #1}}
\def\CMP#1{Comm. Math. Phys.~{\bf #1}}
\def\CNPP#1{Comm. Nucl. Part. Phys.~{\bf #1}}
\def\HPA#1{Helv. Phys. Acta~{\bf #1}}
\def\IJMP#1{Int. J. Mod. Phys.~{\bf #1}}
\def\JP#1{J. Phys.~{\bf #1}}
\def\MPL#1{Mod. Phys. Lett.~{\bf #1}}
\def\NP#1{Nucl. Phys.~{\bf #1}}
\def\NPPS#1{Nucl. Phys. Proc. Suppl.~{\bf #1}}
\def\NC#1{Nuovo Cim.~{\bf #1}}
\def\PH#1{Physica {\bf #1}}
\def\PL#1{Phys. Lett.~{\bf #1}}
\def\PR#1{Phys. Rev.~{\bf #1}}
\def\PRP#1{Phys. Rep.~{\bf #1}}
\def\PRL#1{Phys. Rev. Lett.~{\bf #1}}
\def\PNAS#1{Proc. Nat. Acad. Sc.~{\bf #1}}
\def\PTP#1{Progr. Theor. Phys.~{\bf #1}}
\def\RMP#1{Rev. Mod. Phys.~{\bf #1}}
\def\RNC#1{Riv. Nuo. Cim.~{\bf #1}}
\def\ZP#1{Z. Phys.~{\bf #1}}

\section{Introduction}

Strongly interacting matter is expected to undergo a phase
transition or crossover to a quark--gluon plasma phase
both at high temperature
and at high baryon density.   Since QCD is asymptotically
free, when either the temperature $T$ 
or the Fermi momentum $p_F$ is high the effective coupling 
for typical scattering processes with momentum transfer
of order $T$ or $p_F$ is small, and one therefore expects
a new phase of matter in which color is screened 
rather than confined and chiral symmetry is 
restored.  
The exploration of the phase diagram is of fundamental
interest and has applications in cosmology, in the
astrophysics of neutron stars and in the physics of heavy ion collisions.
The zero density, high temperature axis of the phase diagram
is much better explored than the zero temperature, high
density axis, since lattice Monte Carlo techniques are well-suited
to nonzero temperature but not so well-suited to nonzero
chemical potential.  
Recent work\cite{ARW,shuryak}
suggests a rich phase structure at nonzero density.
First, chiral symmetry restoration proceeds via
a first order phase transition and ordinary nuclear matter may be 
described as being in the mixed phase of this 
transition\cite{ARW}.  Second, above the transition, the quarks
may be weakly coupled but the phase they are in is {\it not} the
trivial one.
Even at weak coupling, any attractive quark-quark interaction
leads to the formation of a condensate of Cooper pairs,
which spontaneously breaks color symmetry.  Color superconductivity
has been studied 
using perturbative methods 
valid at asymptotically high density\cite{bailin},
and in a one-flavor model\cite{iwado}. 
The authors of Refs. \cite{ARW,shuryak}
have used a model in which the interaction
between quarks in two-flavor QCD is modelled by that
induced by instantons\cite{thooft} to estimate
the magnitude of the diquark condensate in the phase
in which chiral symmetry is restored.
We first generalize the variational methods of \cite{ARW}
to a formalism in which we derive a bosonic effective
action for the degrees of freedom which condense.\footnote{Diquark 
condensation has also been sought in the vacuum using similar methods
\cite{dia}.}
For the present investigation, we evaluate the 
effective action using a mean field approximation.
Using the method we describe here,
we can analyze circumstances in which a chiral symmetry
breaking $\langle \bar \psi \psi \rangle$ condensate and
a color breaking $\langle \psi \psi \rangle$ condensate coexist.
Furthermore, we introduce nonzero temperature and
current quark mass.  In this paper, therefore, we are able
to explore the phase diagram as a function
of temperature, density or chemical potential, and quark mass.

The phase diagram which we uncover has striking
qualitative features, several of which we expect to generalize beyond
the model which we consider
and to provide a good guide to the physics of two-flavor QCD.
At low temperatures, chiral symmetry restoration occurs via
a first order phase transition between a phase with low (or zero)
baryon density and a high density phase with a condensate
of quark--quark Cooper pairs in color antitriplet, Lorentz scalar,
isospin singlet states.\footnote{There are also indications\cite{ARW} of
a color ${\bf 6}$, Lorentz axial vector, isospin
singlet condensate which is many orders of magnitude
smaller than the condensates we treat.}
We find color superconductivity 
at temperatures $T<T_c^\Delta$ where $T_c^\Delta$ is
of order tens to almost one hundred  
MeV, and at most a factor of 
four higher if we allow interactions
other than that induced by 
instantons. This suggests that color superconductivity
may arise in heavy ion collisions, in which the necessary
densities are likely accompanied by temperatures of order
$100$ MeV.
The transition we find at $T_c^\Delta$ is second order
for the present model, but the order of this transition 
may change once gauge field fluctuations are taken into
account.
We study the competition between
chiral and superconductor condensates, in particular
when the current quark mass is nonzero and both coexist.
For small quark mass, we find that the color superconductivity
is almost unaffected by the presence of a 
chiral condensate.

In the chiral limit, we 
find a second order 
chiral transition at zero chemical potential and
a first order chiral transition at zero temperature. Consequently, these 
meet at a tricritical point at a particular nonzero
temperature $T_{\rm tc}$ and chemical potential $\mu_{\rm tc}$ as was
first discussed in a different phenomenological model in Ref.\ \cite{fir}.  
We point out that the physics in
the vicinity of the tricritical point is described
by a $\phi^6$ theory, and the transition should therefore
be characterized by the mean field exponents which we
calculate and by logarithmic corrections to scaling\cite{tricrit}.
If two-flavor QCD has a second order transition at
high temperatures and a first order transition at
high densities, then it will have a tricritical point
in this universality class.
Away from the chiral limit, the second 
order chiral transition
turns into a smooth crossover, while the first order
transition remains first order. Of particular
interest is the fact that the tricritical point
becomes an Ising second order transition. This raises the
possibility that even though the pion is massive in nature, 
long correlation lengths may arise in heavy ion 
collisions which traverse the chiral transition
near $T_{\rm tc}$ and $\mu_{\rm tc}$.
Much of the quantitative physics associated with the first order transition
is quite model-dependent.  However, the physics of the region near
the tricritical point is both universal and well-described
in mean field theory.

\section{Models \label{model}}

Our goal is an exploration of
the phase diagram of two flavor QCD
in a context which allows us to describe likely
patterns of symmetry breaking and to make rough quantitative 
estimates. 
As a tractable model, 
we consider a class of fermionic models 
for QCD \cite{NJL} at nonzero temperature
and baryon number density \cite{klevansky}
where the fermions interact via the
instanton-induced interactions between light quarks\cite{thooft}. 
In the Euclidean functional integral for the
effective action, nonzero temperature $T$ is implemented via
antiperiodic boundary conditions for fermionic fields in the 
`time' direction with periodicity $1/T$.
In the presence of a nonvanishing chemical potential $\mu$ 
for (net) quark number, the quadratic part 
of the Euclidean action $S$ in momentum 
space reads\footnote{We use the Euclidean conventions employed
in Ref. \cite{BJW}.} 
\be
S_{0}= T \sum\limits_{n \in \ZZ} \int \frac{d^3 \vec{q}}{(2\pi)^3}
{\bar{\psi}^i}_{\alpha}(q) \left(
\gamma^{\nu}q_{\nu}+i m+ i \gamma^{0} \mu\right){\psi_i}^{\alpha}(q) \, ,
\label{s0}
\ee
with an average current quark mass $m$.
The zeroth component of the momentum $q\equiv (q^0,\vec{q})$ is discrete with 
$q^0(n)=(2n+1)\pi T$
for the fermionic Matsubara modes. Indices $i=1,\ldots,N_f$ and 
$\alpha=1,\ldots,N_c$ denote flavor and color, respectively. 
Spinor indices are suppressed and 
repeated indices are summed. We 
will write expressions in terms of $N_f$ and $N_c$ throughout,
even though we
will only be concerned
with $N_f=2$ and $N_c=3$ in this paper.\footnote{We work 
throughout with the same chemical potential $\mu$
for both up and down quarks.  Color superconductivity
in a system with more down than up quarks, say by about
$15\%$ as in a lead or gold nucleus, has been considered
in Ref. \cite{tsukuba}.  The superconductor 
gap is somewhat reduced, and
this suggests that in a heavy ion collision, the
system can lower its energy
by equalizing the down/up ratio in the densest region of
the plasma in order to minimize the gap.  
This suggests that these dense regions will
expel negative pions toward the periphery.\cite{tsukuba}}

We add a four--fermion interaction to (\ref{s0}) which has
the color, flavor 
and Lorentz structure of the instanton vertex of two--flavor 
QCD\cite{thooft}.  This interaction properly reflects the
chiral symmetry of QCD: axial baryon number is broken, while
chiral $SU(2)_L\times SU(2)_R$ is respected.  Color $SU(3)$ is realized
as a global symmetry. One could add other four-fermion 
interactions that respect the unbroken symmetries of QCD,
including in particular that induced by one gluon exchange.
We will comment on its effects below.
Using the instanton vertex alone is only the simplest way of
breaking all the symmetries QCD breaks while respecting those
it respects.
We note that
with the help of appropriate Fierz transformations the instanton interaction 
can be decomposed into two parts, where
one part contains only color singlet fermion bilinears and the other part 
contains only color $\bar{\bf{3}}$ bilinears: 
$S_I=S_I^{(\bf{1}_c)}+S_I^{(\bar{\bf{3}}_c)}$ with
\ben
S_{I}^{(\bf{1}_c)} &=& G_1 T \sum\limits_{n \in \ZZ} \int
\frac{d^3 \vec{p}}{(2\pi)^3}
\left\{ 
-O_{(\sigma)}[\psi,\bar{\psi};-p] 
 O_{(\sigma)}[\psi,\bar{\psi};p]
-O^a_{(\pi)}[\psi,\bar{\psi};-p] 
 O_{(\pi)a}[\psi,\bar{\psi};p] \right.\nnn && \left. 
\qquad \qquad \qquad \,\,\,\,\,\,\,\,\,\,\,\,
+O_{(\eta^{\prime})}[\psi,\bar{\psi};-p] 
 O_{(\eta^{\prime})}[\psi,\bar{\psi};p]
+O^a_{(a_0)}[\psi,\bar{\psi};-p] 
 O_{(a_0)a}[\psi,\bar{\psi};p]\right\}
\, , \vspace{.3cm} 
\nonumber\\ 
S_{I}^{(\bar{\bf{3}}_c)} &=&  
G_2 T \sum\limits_{n \in \ZZ} \int \frac{d^3 \vec{p}}{(2\pi)^3} \left\{
-{O^{\dagger\alpha}_{(s)}}[\bar{\psi};p]
  O_{(s)\alpha}[\psi;p]
+{O^{\dagger\alpha}_{(p)}}[\bar{\psi};p]
  O_{(p)\alpha}[\psi;p] \right\}\ .
\label{si3}
\een
Here $p\equiv (2n\pi T,\vec{p})$ due to the bosonic nature
of the fermion bilinears, and $^\dagger$ denotes the
operation of Euclidean reflection.
The bilinears $O_{(\sigma)}$, $O^a_{(\pi)}$,
$O_{(\eta^{\prime})}$ and $O^a_{(a_0)}$ with $a=1,2,3$ carry the quantum 
numbers associated with the scalar isosinglet ($\sigma$), the 
pseudo-scalar isotriplet ($\pi$), the pseudo-scalar 
isosinglet ($\eta^{\prime}$)
and the scalar isotriplet ($a_0$), respectively. Similarly, 
the bilinear $O_{(s)}^{\alpha}$ ($O_{(p)}^{\alpha}$), 
with the color index $\alpha=1,2$ or $3$, 
carries the quantum numbers of the color 
antitriplet scalar (pseudo--scalar) diquark. 
(See (\ref{osigma}) below for explicit expressions.)
The instanton interaction introduces only one coupling.
If we take this to be $G_1$, then $G_2=G_1/(N_c-1)$ in
(\ref{si3}).  We have generalized the interaction to 
allow $G_1$ and $G_2$ to take on independent values,
although we will always assume that they have the
same sign, as for the instanton interaction.
We discuss choices of $G_1$ and $G_2$ at greater length below.
We note from the signs in $S_I$ that if we choose 
the sign of $G_1$ such that the interaction in
the $\sigma$ channel is attractive, so that chiral symmetry
breaking is favored, we may expect condensation in
the $\pi$ and scalar diquark channels also.
In the chiral limit, one can always
make a rotation such that there is no $\pi$ condensate.
A condensate in the pseudoscalar diquark channel would break parity
spontaneously, but this seems not to be favored by the 
interaction (\ref{si3}).
We will use the model to explore condensates in
the $\sigma$ and scalar diquark channels.

In order to mimic 
the effects of asymptotic freedom, the interaction has to decrease with 
increasing momentum. We follow Ref. \cite{ARW} 
and implement this via a form factor $F(\vec{q})$
in momentum space 
for the three--momenta of each of the fermions,  
with $F(0)=1$ and $F(\vec{q}) \ll 1$ for 
${\vec{q}}^{\,2} \gg {\Lambda}^2$. 
Here $\Lambda$ is some effective
QCD cutoff scale which one might anticipate 
to be in the range $1/2 - 1$ GeV. 
With form factors included, the fermion bilinears in the  
above interaction read
\begin{eqnarray}
O_{(\sigma)}[\psi,\bar{\psi};p] &=& -i T \sum\limits_{n \in \ZZ} \int
\frac{d^3 \vec{q}}{(2\pi)^3}
F(\vec{q}) F(\vec{p}-\vec{q}) {\bar{\psi}}^{i}_{\alpha}(-q) 
{\psi_i}^{\alpha}(p-q) \ , \nonumber \\
O_{(s)}^{\alpha}[\psi;p] &=& T \sum\limits_{n \in \ZZ} \int
\frac{d^3 \vec{q}}{(2\pi)^3}
F(-\vec{q}) F(-\vec{p}+\vec{q})\, {{(\psi^T)}^i}_{\beta}(-p+q)\, C \gamma^5\, 
\epsilon^{\alpha\beta\gamma}\epsilon_{ij}\,
{\psi^j}_\gamma(-q) \ ,
\label{osigma}
\end{eqnarray}
for the $\sigma$ and for the scalar diquark
and similarly for the other
bilinears. $C$ denotes the charge conjugation matrix.

\section{Thermodynamic Potential}

The fermionic effective action $\Gamma$ (the generating functional
for $1PI$ Green functions) determines the field equations that
contain all quantum effects. In thermal and chemical equilibrium,
$\Gamma$ depends on the temperature $T$ and chemical
potential $\mu$. Here we are interested in a computation of the
effective action that determines the field equations for the
expectation values of the fermion bilinears 
$O_{(\sigma)}[\psi,\bar{\psi};x]$ and $O_{(s)}^{\alpha}[\psi;x]$
whose Fourier components are given in (\ref{osigma}). We denote
these expectation values by $\langle\bar{\psi}\psi\rangle$ 
and $\langle \psi\psi\rangle$, respectively. The chiral condensate
$\langle\bar{\psi}\psi\rangle$ is an order parameter for chiral
symmetry breaking for vanishing current quark mass $m$. 
For nonzero $\langle \psi\psi\rangle$,
the diquark bilinear has a
color index $\alpha$ 
which chooses a direction in color space, thus breaking 
color symmetry $SU(3) \rightarrow SU(2)$.\footnote{
The $U(1)$ of electromagnetism is spontaneously broken but there is
a linear combination of electric charge and color hypercharge under
which the condensate is neutral which therefore generates an unbroken
$U(1)$ gauge symmetry.} In the following we compute the effective
action as a function of the two (space dependent) order 
parameters.\footnote{We note that the diquark condensate
breaks the gauge symmetry and is therefore 
not a gauge invariant order parameter. In this respect the situation
resembles that in the electroweak sector.
As in that case, there is the possibility that a transition associated
with a rapid change in physical properties
may proceed through a smooth crossover without any 
thermodynamic singularity.}
Its behavior for constant order parameters yields the effective
potential which corresponds at its extrema to the thermodynamic potential
and encodes the equation of state.   

For our purposes, it is advantageous
to introduce bosonic collective fields into the functional
integral which defines the fermionic effective action \cite{bos}. 
These collective
fields carry the quantum numbers of the fermion bilinears 
appearing in (\ref{si3}). 
We employ a Hubbard-Stratonovich transformation,
in which the collective fields are introduced into the functional
integral by inserting the identities
\ben
1 &=& N\! \int \!\Dc \phi \exp\!\left\{\! -T \sum\limits_{n \in \ZZ} \!
\int \frac{d^3 \vec{p}}{(2\pi)^3} \left(\! \frac{\phi(-p)}{2}
-O_{(\sigma)}[\psi,\bar{\psi};-p] 
G_1\! \right)
\frac{1}{G_1}
\left(\! \frac{\phi(p)}{2}
- G_1 O_{(\sigma)}[\psi,\bar{\psi};p]\! \right)\right\} ,
\nnn
1 &=& \tilde{N}\! \int \! \Dc \Delta^{*} \Dc \Delta 
\exp\! \left\{\! -T \sum\limits_{n \in \ZZ} \!
\int \frac{d^3 \vec{p}}{(2\pi)^3}\left(\! \frac{\Delta^{*\alpha}(p)}{2}-
O_{(s)}^{\dagger\alpha}[\bar{\psi};p] 
G_2\! \right)
\frac{1}{G_2} 
\left(\! \frac{\Delta_\alpha(p)}{2}- G_2 O_{(s) \alpha}[\psi;p]\! 
\right)\right\}
\label{hub} \nonumber \\
\een
and similar identities for the other bilinears.
(Here $N$ and $\tilde{N}$ are field independent normalization factors.)
The terms quadratic in the fermion bilinears in (\ref{hub}) cancel
the original four--fermion interaction (\ref{si3}).
As a result, the four--fermion interaction is cast into a
Yukawa interaction between fermions and collective fields and
a mass term for the collective fields. The
fermionic fields then appear only quadratically and can be 
integrated out exactly, leaving an effective
action for the bosonic collective fields alone. 
We apply a saddle point expansion
to the resulting effective action. In particular, standard mean field results 
correspond to the lowest order in this expansion. Having done the 
Gaussian integral for the fermions, the saddle point effective action 
reads  
\ben
\Gamma [\phi,\Delta,\Delta^*;T,\mu] &=& - \tr \ln H^{21} - \frac{1}{2} 
\tr \ln \left[{\bf 1}-(H^{21})^{-1}H^{22}(H^{12})^{-1}H^{11}\right] \nnn
&&+T \sum\limits_{n \in \ZZ} \int
\frac{d^3 \vec{p}}{(2\pi)^3} \left[ \frac{1}{4G_1} \phi(-p) \phi(p)
+ \frac{1}{4G_2} \Delta^{*}_{\alpha}(p)\Delta^{\alpha}(p) \right] 
\label{saddel}
\een
where, for constant fields $\phi(x)=\phi$ and (taking the color index
$\alpha=3$)
$\Delta_{3}(x)=\Delta_{3}^{*}(x)=\Delta$ and $\Delta_1=\Delta_2=0$, 
the $H^{ab}$ are 
diagonal in momentum space:
\ben
(H^{11})^{\alpha\beta}_{ij}(q,q^{\prime}) &=& - C \gamma_5 \epsilon^{\alpha\beta 3}
\epsilon_{ij} F(\vec{q})^2 \Delta 
\delta^3(\vec{q}-\vec{q}^{\,\, \prime}) \delta_{n,n^{\prime}}/2\pi T
= -(H^{22\dagger})^{\alpha\beta}_{ij}(q,q^{\prime})
\ , \nnn
(H^{12})^{\alpha\beta}_{ij}(q,q^{\prime}) &=& 
- [C(-\gamma^{\mu}q_{\mu}+i\gamma^0\mu)C
+i(m+F(\vec{q})^2\phi)]\delta^{\alpha\beta}\delta_{ij}
\delta^3(\vec{q}-\vec{q}^{\,\, \prime}) \delta_{n,n^{\prime}}/2\pi T\ , \nnn
(H^{21})^{\alpha\beta}_{ij}(q,q^{\prime}) &=& 
[(\gamma^{\mu}q_{\mu}+i\gamma^0\mu)
+i(m+F(\vec{q})^2\phi)]\delta^{\alpha\beta}\delta_{ij}
\delta^3(\vec{q}-\vec{q}^{\,\, \prime}) \delta_{n,n^{\prime}}/2\pi T\ .
\label{Hab}
\een
The integration
over the fermions gives rise to a determinant which has been rewritten
as the trace of a logarithm in (\ref{saddel}). 
The trace involves the momentum integration as well
as the summation over flavor, color and spinor indices. 
In principle, the saddle point effective action depends 
on the expectation values of all fermion bilinears occuring in (\ref{si3}).
The corresponding collective fields would modify the $H^{ab}$ of 
(\ref{Hab}) in a straightforward way and for each field a quadratic,
mass like, term appears in (\ref{saddel}). As discussed
in the previous section, these
collective fields are not expected to condense. They therefore only
contribute to higher order terms in the saddle point expansion.   

In this work, we restrict 
ourselves to the mean field approximation though 
the approach can be extended to
take into account fluctuations around the saddle point. The mean field 
approximation corresponds to the summation of an infinite class of 
diagrams, but it is not a systematic expansion in powers of a small 
parameter.
A nonperturbative approach beyond mean field may be
performed along the lines of Ref.\ \cite{BJW}. There, the two flavor
effective potential has been computed
at zero baryon density 
using renormalization group methods. 
In particular, this approach allows a precise treatment
of the effective potential in the vicinity of second
order\cite{BTW,BJW} or weak first order\cite{BW} transitions,
both of which we will encounter in Section V.
We leave the extension of these methods
to nonzero baryon density for future work, as our goal here is to use
mean field results as a qualitative guide in exploring the phase diagram.

We evaluate the effective action in the saddle point
approximation for constant field configurations. This yields the 
effective potential or thermodynamic potential 
$\Omega=\Gamma \, T/V$.
A convenient way to extract the effective potential 
from (\ref{saddel}) is to consider the derivatives of $\Gamma$ with 
respect to $\phi$ and $\Delta$. After performing the traces and,
in particular, the sums over Matsubara frequencies, $\Omega$
can be obtained by integration. We find
\ben
\Omega(\phi,\Delta;\mu,T) &=& 
\frac{1}{4 G_1} \phi^2 + \frac{1}{4 G_2} \Delta^2 \nnn \qquad
&-& 2 N_F \int\limits_0^{\infty} 
\frac{q^2 dq}{2 \pi^2} \Biggr\{ 
(N_c-2) \Big\{ E_{\phi} 
+T \ln \Big(1+\exp\Big[-(E_{\phi}-\mu)/T\Big]\Big) \nnn \qquad
&\ &\ \ \ \ \ \ \ \ \ \ \ \ \ \ \ \ \ \ \ \ \ \ \ \ \ \ \ \ \ \ \ \ \   
\ \ +T \ln \Big(1+\exp\Big[-(E_{\phi}+\mu)/T\Big]\Big)
\Big\} \nnn 
&~& \nnn
&+&(E_{\phi}-\mu) \sqrt{1+F^4 \Delta^2/(E_{\phi}-\mu)^2}
+(E_{\phi}+\mu) \sqrt{1+F^4 \Delta^2/(E_{\phi}+\mu)^2} \nnn 
&+&2 T \ln\left(1+\exp\left[-(E_{\phi}-\mu)
\sqrt{1+F^4 \Delta^2/(E_{\phi}-\mu)^2}/T\right]\right)\nnn  
&+&2 T \ln\left(1+\exp\left[-(E_{\phi}+\mu)
\sqrt{1+F^4 \Delta^2/(E_{\phi}+\mu)^2}/T\right]\right)\Biggl\}
\ +\  {\rm constant}
\label{potential}
\een
where $F(\vec{q})$ is the form factor and
\be
E_{\phi}(\vec{q})\equiv\sqrt{{\vec{q}\, }^2+(m+F(\vec{q})^2 \phi)^2} \,\, .
\ee
The constant in (\ref{potential}) does not depend on $\mu$ and $T$
and is chosen such that the pressure of the physical vacuum is 
zero. 
Extremizing $\Omega$ leads to coupled gap equations for 
the two order parameters, whose solutions we call
$\phi_0$ and $\Delta_0$.  That is,
\begin{equation}
\frac{\pa \Omega}{\partial\phi}\Bigg|_{\phi=\phi_0; \Delta=\Delta_0}
=\frac{\pa \Omega}{\partial\Delta}\Bigg|_{\phi=\phi_0; \Delta=\Delta_0}
=0\ 
\label{gapeqs}
\end{equation}
where $\phi_0$ and $\Delta_0$ are related to the chiral condensate
and the condensate of Cooper pairs by 
\be
\phi_0 = 2 G_1 \Big{\langle}\bar{\psi}\psi \Big{\rangle}\, , \qquad
\Delta_0 = 2 G_2 \Big{\langle} \psi\psi \Big{\rangle} .
\ee
At its extrema, the potential
is related to the energy density $\epsilon$, 
the entropy density $s$, the quark number density $n$ and the pressure $P$ 
by
\be
\Omega(\phi_0,\Delta_0;\mu,T) = \epsilon -T s - \mu n = -P\ .
\ee
Unless otherwise
stated, the results we quote in our exploration of
the phase diagram are obtained
using the smooth form factor 
\begin{equation}
F(\vec{q})=\Lambda^2/({\vec{q}}^{\,2}+\Lambda^2)
\end{equation}
with $\Lambda=0.8$ GeV 
and with $G_1\Lambda^2=6.47$ fixed by requiring
$\phi_0=\phi_0^{\rm vac}=0.4$ GeV at $\mu=T=0$ in order to
obtain a reasonable, albeit qualitative, phenomenology. 
For most of the discussion we will consider $G_2=3G_1/4$,
which will be motivated at length below. We note that the 
qualitative features which we address do not depend
on this specific choice and we discuss the effects
of different choices below.

Although we describe our results in the next sections, it
is worth pausing here to emphasize how we 
use $\Omega$.  
As $T$ or $\mu$ or $m$ changes, 
$\Omega$  can have several local minima in
the $(\phi,\Delta)$ plane.  The lowest minimum
describes the lowest free energy state and is favored.
As an example, which we discuss at length in the next
section, in Figure 1 we plot $\Omega(\phi,\Delta)$ 
at $T=0$ and $\mu=0.292$ GeV. 
One observes two degenerate minima corresponding to 
a first order phase
transition at which two phases 
have equal pressure
and can coexist. 
We discuss the 
symmetry breaking patterns, their effects and
the varied phase transitions which occur as a function
of temperature and chemical potential or density 
in Sections IV--VI.
\begin{figure}[t]
  \unitlength1.0cm
  \begin{center}
  \begin{picture}(13.0,8.0)
  \put(-1.0,-1.0){
  \epsfysize=8.cm
  \epsfbox[80 480 460 720]{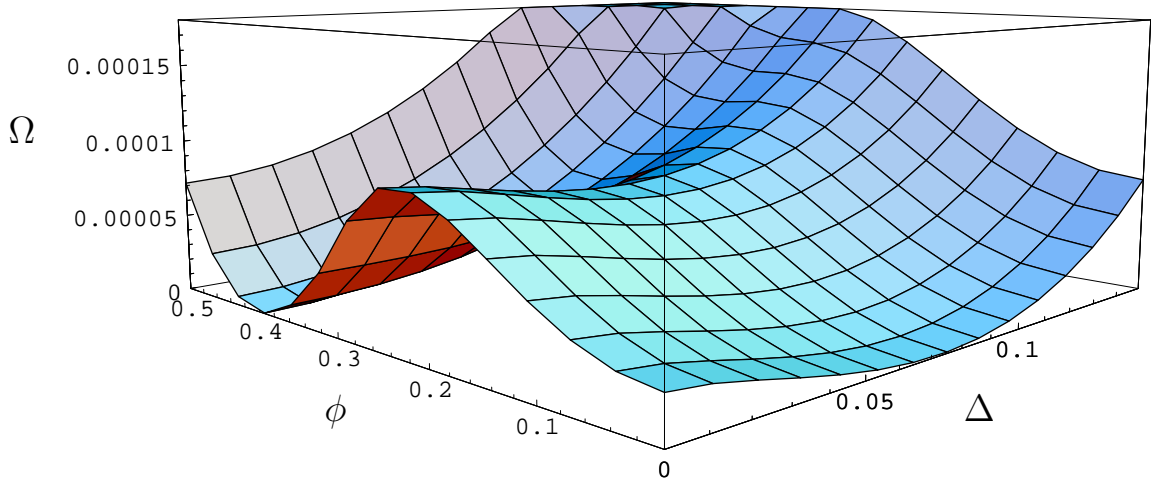}}
  \put(-1.2,4.9){\bf\large $\ds{\Omega}$}
  \put(3,1.2){\bf\large $\phi$}
  \put(11.5,1.2){\bf\large $\Delta$}
  \end{picture}
  \end{center}
\caption{The thermodynamic potential $\Omega$ (in GeV$^4$) as a function
of $\phi$ and $\Delta$ at $T=0$ and  
$\mu=0.292$ GeV.  The two degenerate minima 
have $\phi=\phi_0^{\rm vac}=0.4$ GeV,
$\Delta=0$ 
and $\phi=0$, $\Delta=0.072$ GeV.}
\end{figure}

Before proceeding, we return to our postponed
discussion of the appropriate choice
for $G_2/G_1$, having fixed $G_1$ phenomenologically.
If we take $G_2$ to be independent
of $G_1$, there is a phenomenological upper bound which $G_2$ must
satisfy.  It is only for $G_2 < 2 G_1$ that the
minimum of the potential at $\mu=T=0$ has $\Delta=0$.
For larger $G_2$, color would be spontaneously broken
in the vacuum (see also \cite{dia}). (As $G_2$ increases above $2 G_1$, 
the vacuum state moves continuously in the 
$(\phi,\Delta)$ plane from the $\Delta=0$
axis to the $\phi=0$ axis.)  
It is not unreasonable to leave $G_2$ as a free parameter satisfying
$G_2<2G_1$ for the following reason.  One could
extend our model by introducing four fermion
interactions different from (\ref{si3}) with new coupling
constants, as long as the correct symmetries are respected.
These could again be Fierz transformed into products
of fermion bilinears and treated as described above.
Assuming that condensation occurs only in 
the $\sigma$ and scalar diquark
channels, 
we would end up with a mean field thermodynamic potential
precisely of the form (\ref{potential}), but with 
$G_1$ and $G_2$ replaced by rescaled couplings $G_1'$ and $G_2'$.
However, there is a caveat if 
a vector interaction of the form 
$(\bar \psi \gamma^\mu \psi)^2$ is present. As with any four fermion 
interaction, this can rescale $G_1$ and $G_2$. However, because 
$\langle \bar \psi \gamma^0 \psi \rangle =n \neq 0$, there is an additional 
qualitative effect. In mean field, 
the vector interaction translates into an $n^2$ term added to 
(\ref{potential}) and an effective shift
in the chemical potential proportional to the density
$n$ \cite{klevansky}.
We will address the consequences of such an interaction in 
Section IV.

Suppose that, instead of leaving
$G_2/G_1$ unspecified, we wish to treat the instanton interaction
for which $G_2=G_1/(N_c-1)$.
If we could analyze the model without approximation,
we would use this ratio.
However, 
the decomposition of the instanton interaction in (\ref{si3}) is,
of course, not 
unique.  It can easily be Fierz transformed at the cost
of generating products of other bilinears which are, 
for example, in color octet or ${\bf 6}$ representations.
Though this is of no concern
were we to treat the model exactly, 
it introduces an ambiguity in the mean field treatment we employ.
This is because in addition to generating new bilinears, the
Fierz transformation changes the couplings in front
of the bilinears in (\ref{si3}). In complete analogy
to the above discussion of new interactions, this would change
the result (\ref{potential}) for
$\Omega$ by a rescaling of $G_1$ and $G_2$.
This ambiguity is a familiar one in studies of
chiral symmetry breaking in NJL models, where it
is well known that a mean field calculation of the
chiral condensate reproduces the results of 
summing direct and exchange diagrams if one adds
to the original interaction a suitably Fierz transformed
interaction \cite{klevansky}.  Applied to (\ref{si3}), 
this scales $G_2$ to zero, 
removing the diquark bilinears, and introduces color octet terms.
This procedure is correct only if no diquark condensate forms,
as is of course the case at low densities.
Because we are considering the
possibility of diquark condensation in addition to chiral condensation,
we must then Fierz transform the interaction (including 
the color octet terms) yet again,
this time into a form with $G_1$ scaled to zero,
containing no quark--anti-quark bilinears.
Adding this to the interaction without diquark bilinears,\footnote{This 
is equivalent
to the observation that whereas chiral condensation uses
the instanton vertex to turn a $\bar u_L u_R$ into a $\bar d_R d_L$,
condensation in the scalar diquark channel uses it to 
turn $u_L d_L $ into $u_R d_R$, and so one must add
the exchange interaction in this sense also.} we obtain
(\ref{si3}) with 
\begin{equation}
G_2 = G_1\,\frac{N_c}{2N_c-2}\ ,
\label{g2g1}
\end{equation}
and with color octet and ${\bf 6}$ terms which we now discard
because they are assumed not to condense.
This procedure yields
gap equations
which agree, when either $\phi$ or $\Delta$
is set to zero,
with those 
obtained previously\cite{ARW} 
via extremizing a variational wave function 
and we use the ratio (\ref{g2g1}) in all the 
results we quote unless stated otherwise.
The variational method has the advantage of
avoiding these ambiguities in the mean field approach we are using here. 
Our present
analysis has the virtue of
being tractable at nonzero temperature, for nonzero quark
mass and, perhaps most important, in the presence of simultaneous
condensation in the $\sigma$ and scalar diquark channels.

\section{Zero Temperature Physics}

We begin our analysis
of the thermodynamic potential (\ref{potential})
at zero temperature and in
the chiral limit.  $\Omega(\phi,\Delta;\mu,T)$ is a function
of two order parameters.  As $\mu$ changes, $\Omega$
can have several local minima in the $(\phi,\Delta)$
plane.  An example has been given in Figure 1, where
for $\mu=\mu_0=0.292$ GeV the potential has two degenerate
minima.  To see how the potential changes with $\mu$,
in Figure 2 we display sections of the $(\phi,\Delta)$
plane for $\Delta=0$ and $\phi=0$ respectively.
Let us first focus on the left panel.  At $\mu=0$,
$\Omega$ has a minimum
at $\phi=0.4$ GeV.  As $\mu$ is increased, $\Omega$ at 
this minimum is unchanged.  Since 
\begin{figure}[t]
  \unitlength1.0cm
  \begin{center}
  \begin{picture}(13.0,6.)
  \put(-2.0,0.0){
  \epsfxsize=8.2cm
  \epsfysize=6.cm
  \epsfbox[80 480 460 720]{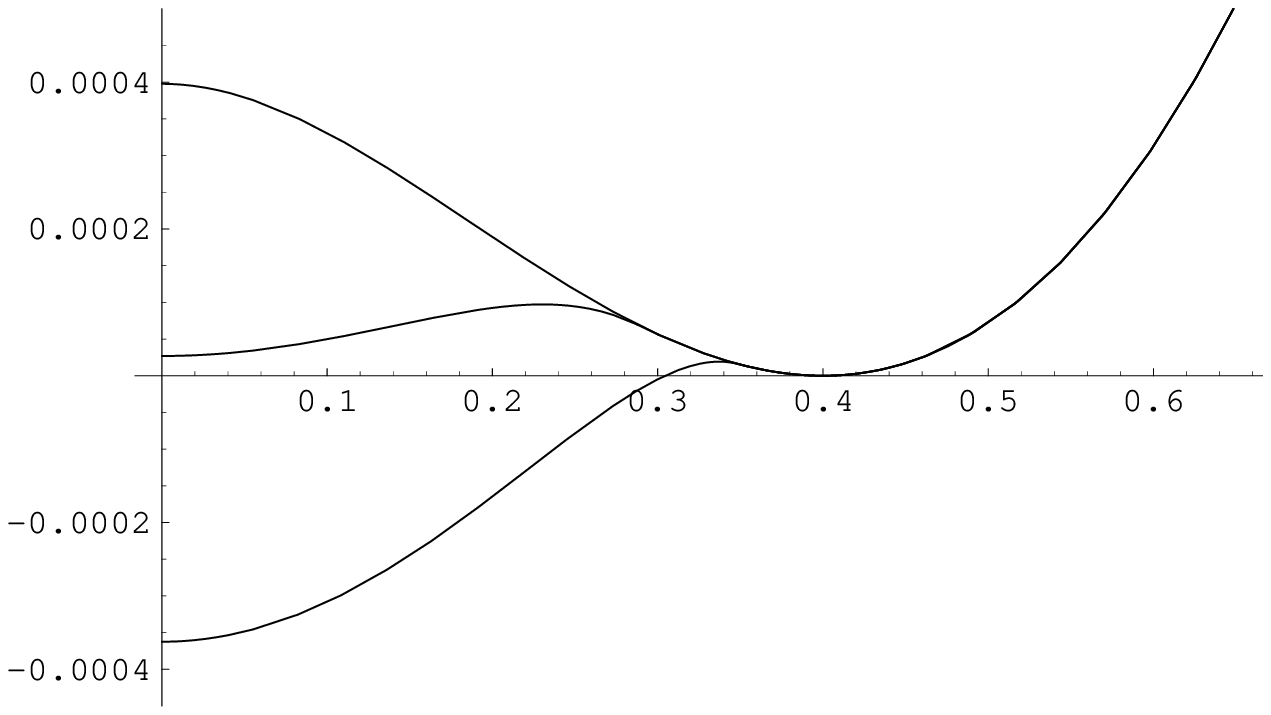}}
  \put(6.2,0.0){
  \epsfxsize=8.2cm
  \epsfysize=6.cm
  \epsfbox[80 480 460 720]{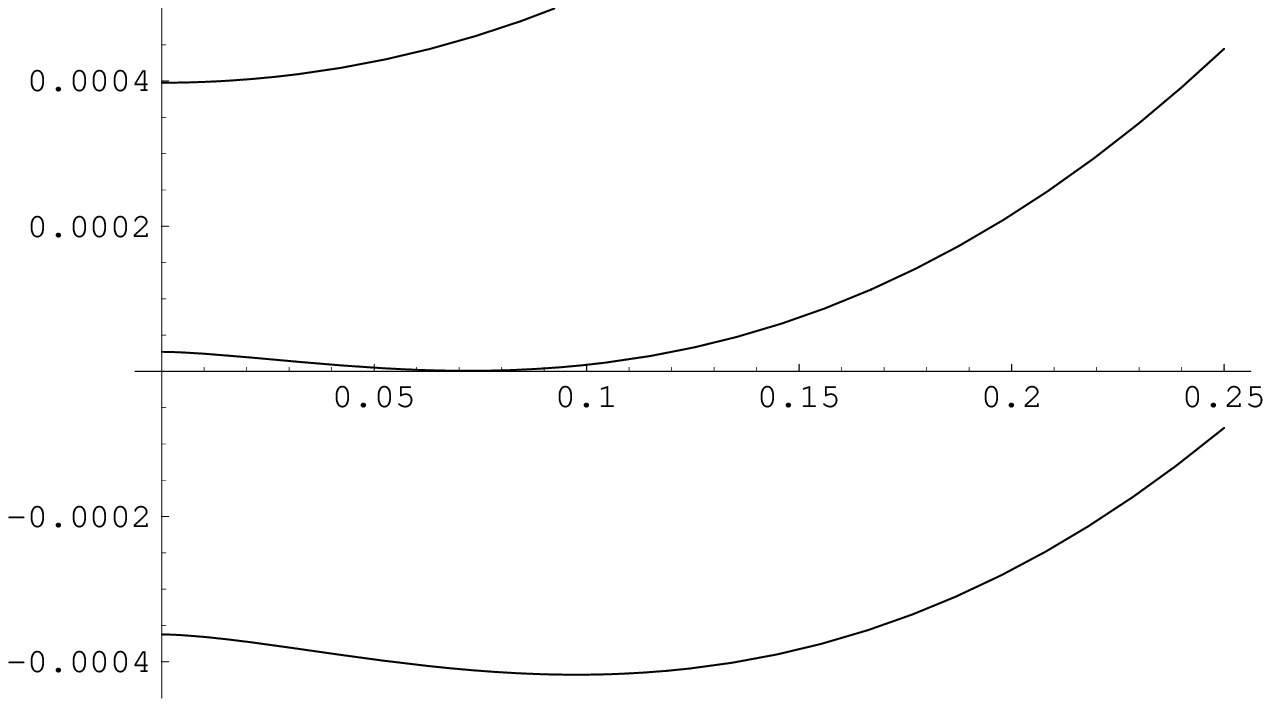}}
  \put(-1.2,5.5){\bf $\Omega$}
  \put(5.9,3.3){\bf $\phi$}
  \put(7.0,5.5){\bf $\Omega$}
  \put(14,3.3){\bf $\Delta$}
  \end{picture}
  \end{center}
\caption{The zero temperature thermodynamic potential 
$\Omega$ (in GeV$^4$) as a function
of $\phi$ at $\Delta=0$ (left panel) and 
as a function of $\Delta$ at $\phi=0$ (right panel)
for several chemical potentials. The curves 
correspond to (top to bottom) $\mu=0,0.292,0.35$ GeV. 
The curves at $\mu=\mu_0=0.292$ GeV are sections of Figure 1.}
\end{figure}
\begin{equation}
n=-\frac{\partial \Omega}{\partial \mu}\ \ {\rm at~fixed~} T, \phi=\phi_0, 
\Delta=\Delta_0,
\label{ndef}
\end{equation}
we conclude that the phase described by this minimum has baryon
density zero, as we expect for the vacuum.  As $\mu$ is
increased sufficiently, a second local minimum of $\Omega$ develops
with $\phi=0$ and $\Delta\neq 0$.
The system remains in the vacuum state with
density zero and $\phi_0$ unchanged until the chemical potential
is increased to $\mu=\mu_0$ where both minima become
degenerate with $\Omega=-P=0$ as seen in Figure 1. 
There is now a zero pressure
chirally symmetric color superconductor
phase whose density we denote $n_0$ which has
the same $\mu$ and $P$ as the vacuum phase.
Before $\mu$ can be increased any further, the system
must undergo a first order phase transition during which
the density increases from zero to $n_0$. At all
intermediate densities, the system is in a mixed phase
in which there are regions with density $n_0$ and 
regions of vacuum.  Once the transition is complete,
$\mu$ can increase further, and the lowest $\Omega$
configuration becomes one with $\Omega<0$, that is with
positive pressure, in which $n>n_0$.  
\begin{figure}[t]
  \unitlength1.0cm
  \begin{center}
  \begin{picture}(13.0,11.0)
  \put(-2.0,5.7){
  \epsfysize=5.2cm
  \epsfbox[80 480 460 720]{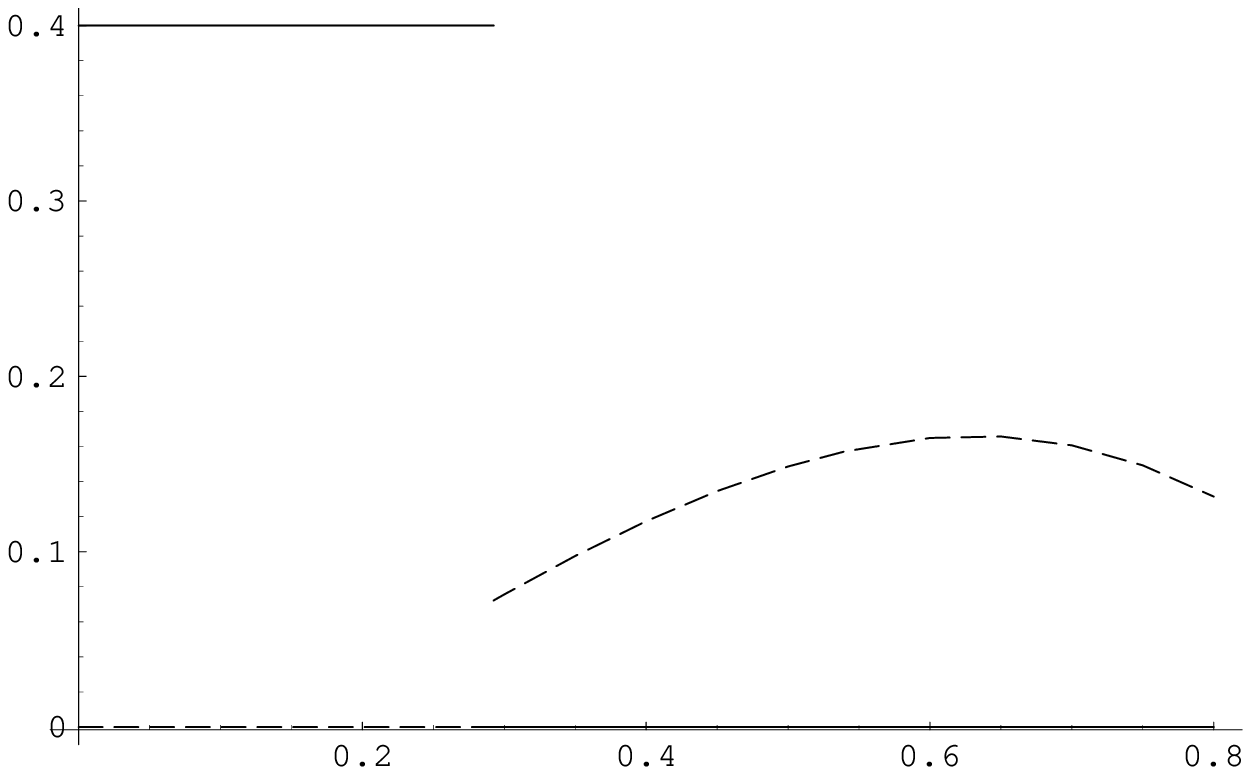}}
  \put(6.1,5.7){
  \epsfysize=5.21cm
  \epsfbox[80 480 460 720]{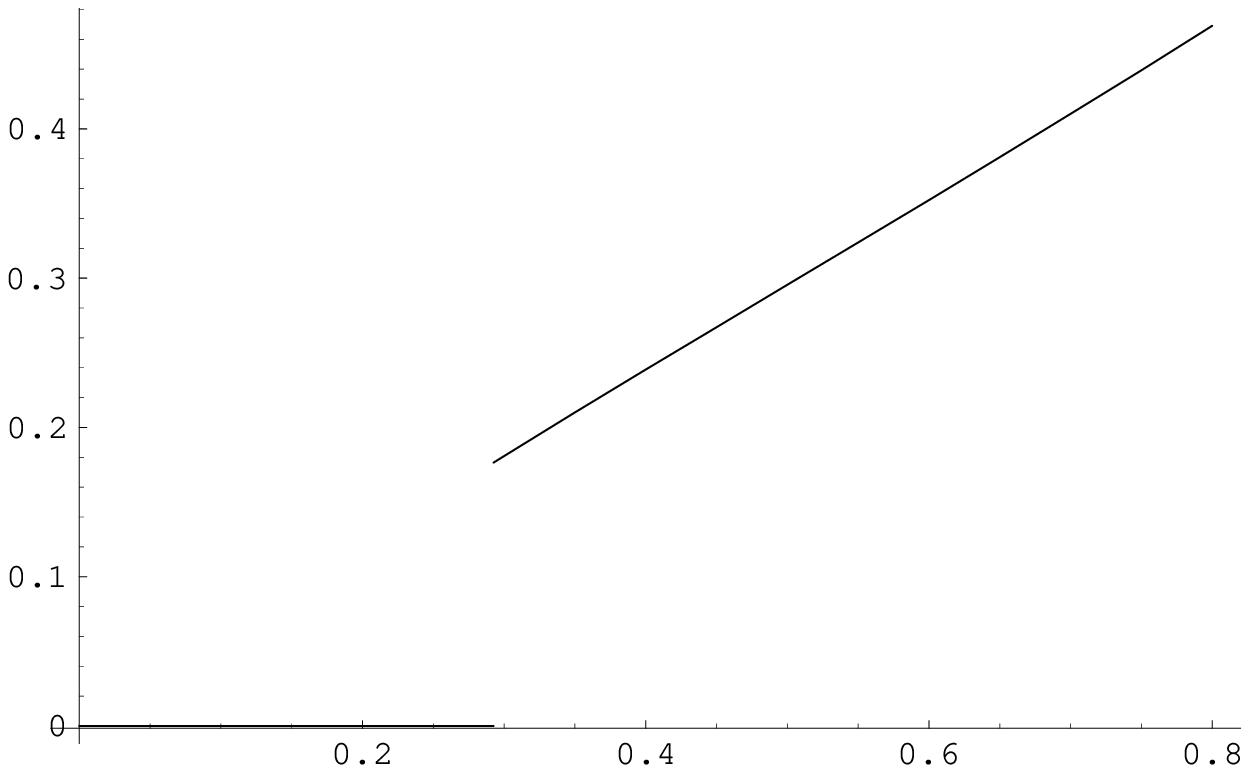}}
  \put(-2.0,0.2){
  \epsfysize=5.2cm
  \epsfbox[80 480 460 720]{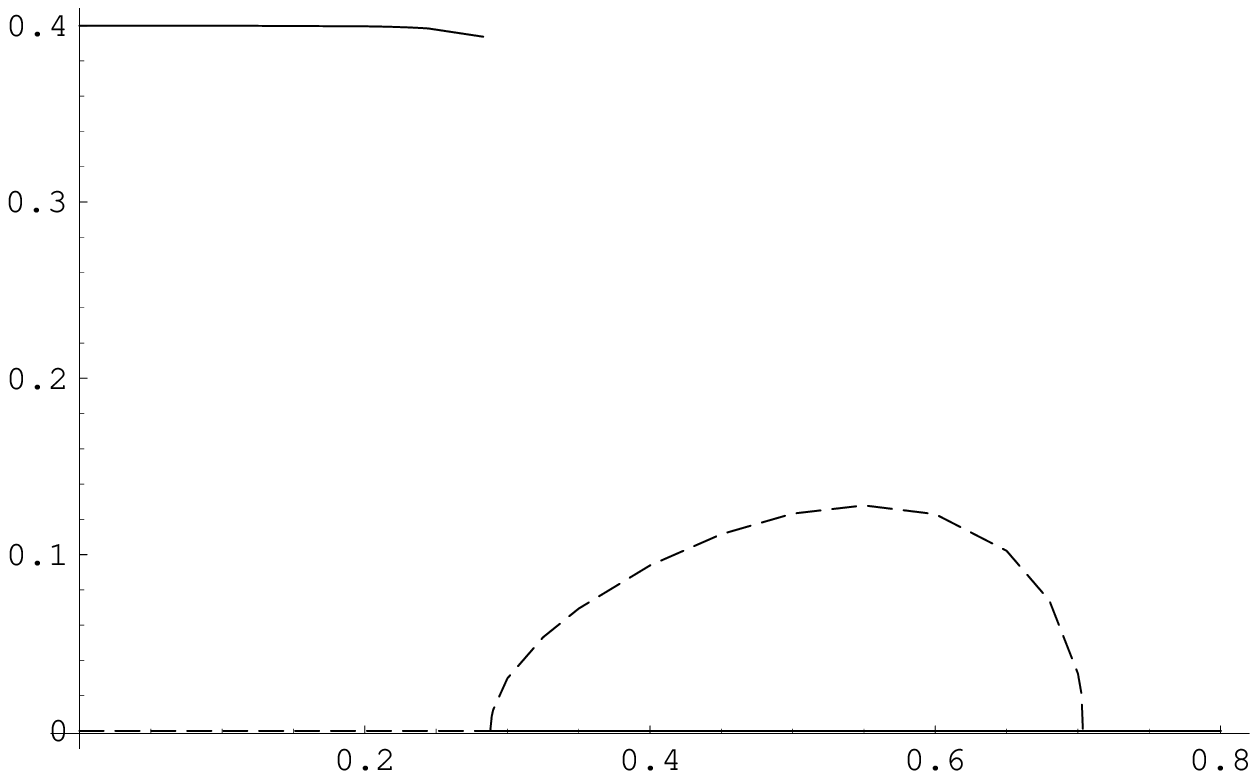}}
  \put(6.1,0.2){
  \epsfysize=5.2cm
  \epsfbox[80 480 460 720]{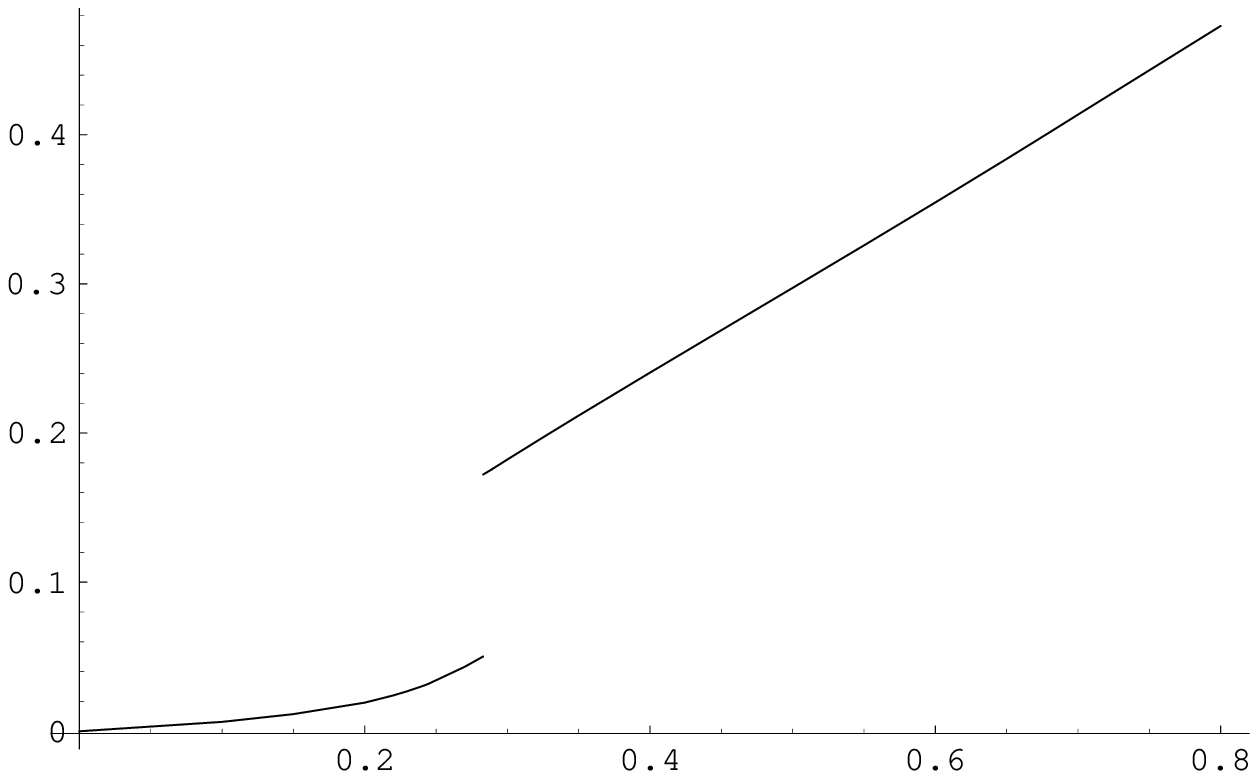}}
  \put(7.5,4.9){\bf $n^{1/3}$}
  \put(0,4.8){\bf $\phi_0$}
  \put(0,10.3){\bf $\phi_0$}
  \put(7.5,10.4){\bf $n^{1/3}$}
  \put(3.5,8.3){\bf $\Delta_0$}
  \put(3.5,2.45){\bf $\Delta_0$}
  \put(2.5,0.2){\bf $\mu$}
  \put(10.6,0.2){\bf $\mu$}
  \end{picture}
  \end{center}
\caption{Left panels show
$\phi_0$ and $\Delta_0$ (namely $\phi$ and $\Delta$
at the global minimum of $\Omega$) as functions of $\mu$
for $T=0$ (above) and $T=0.03$ GeV (below). 
The right panels show the quark number density $n$
for the same parameters.  
We describe the results at nonzero temperature in
the next section.}
\end{figure}
In the top panels of Figure 3, we show the behavior of
the global minimum $(\phi_0,\Delta_0)$ of $\Omega$ and
the density as a function of $\mu$.
We have verified that there are no other minima, for
example with both $\phi$ and $\Delta$ nonzero, as long as we stay
in the chiral limit.
The locations of the extrema of the curves in Figure 2 
agree with those obtained in Ref. \cite{ARW}
and described also in Ref. \cite{ykis97}.
Treating the full problem, one observes that the
first order phase transition does in fact occur between the two 
minima in Figure 1, at $\mu=\mu_0=0.292$ GeV.

There is an interesting interpretation
of this zero temperature first order 
phase transition\cite{ARW,ykis97,buballa}.  
For reasonable choices of parameters, 
$n_0$ is greater
than nuclear matter density and comparable to the baryon density
in a {\it nucleon}.\footnote{If we identify $n_0$ with
the quark number density in
a nucleon and take this to be three times 
that in nuclear
matter, we should require $n_0^{1/3}=0.22$ GeV.  
The choice of parameters we are using for illustration
yields a smaller value $n_0^{1/3}=0.18$ GeV in the chiral limit, 
as seen in Figure 3. We shall see in Section VI that we obtain
$n_0^{1/3}=0.22$ GeV for a quark mass of $10$ MeV.}
Ordinary
nuclear matter is then in the mixed phase of this transition and
consists of nucleon sized droplets within which $n=n_0$ and
$\phi=0$, surrounded by regions of vacuum.  Although it
is nice to see that the model forces quarks to be located within
droplets within which they have zero mass, the model
must obviously be extended in several respects. First, a short range
repulsion between droplets must be included, so that small
droplets are favored over bigger ones \cite{ykis97}. Second, 
color must be gauged if the model is
to yield droplets which are color singlets.  
It is crucial to this interpretation that $\mu_0<\phi_0^{\rm vac}$.
In the chiral limit, $\phi_0$ corresponds to the
constituent quark mass and therefore if $\mu$ is 
increased above $\phi_0^{\rm vac}$ the (formerly) vacuum phase becomes
populated with a gas of constituent quarks.  If $\mu_0 < \phi_0^{\rm vac}$,
the first order transition to the high density phase
which occurs at $\mu=\mu_0$ preempts this.
Within the present model, $\mu_0<\phi_0^{\rm vac}$ only for
$\Lambda<2.2$ GeV.  Furthermore, adding a four fermion interaction
to model one gluon exchange introduces an effective shift in $\mu$ as
discussed in Section III, and this increases the value 
of $\mu_0$\cite{klevansky}.  With vector interactions included,
one can construct models with reasonable values
of a suitably defined $\Lambda$ in which 
$\mu_0>\phi_0^{\rm vac}$ \cite{klevansky}.
In such a model, the first order transition occurs between
a gas of constituent quarks with a nonzero density $n_-$
and a gas of massless quarks with a density $n_+>n_-$,
and 
$n_-$ can be greater than nuclear matter density
for reasonable choices of parameters.  
If one then gauges color, the gas of constituent quarks
which the model describes at densities below $n_-$ 
is expected to become a gas of baryons.  
We now turn to nonzero temperature  and one of the
things which we will see is that 
models with $n_-=0$ and with $n_->0$ 
are less different than they appear to be at zero
temperature.

\section{The Phase Diagram}

\begin{figure}[t]
  \unitlength1.0cm
  \begin{center}
  \begin{picture}(13.0,14.3)
  \put(0.0,7.1){
  \epsfysize=7.8cm
  \epsfbox[80 480 460 720]{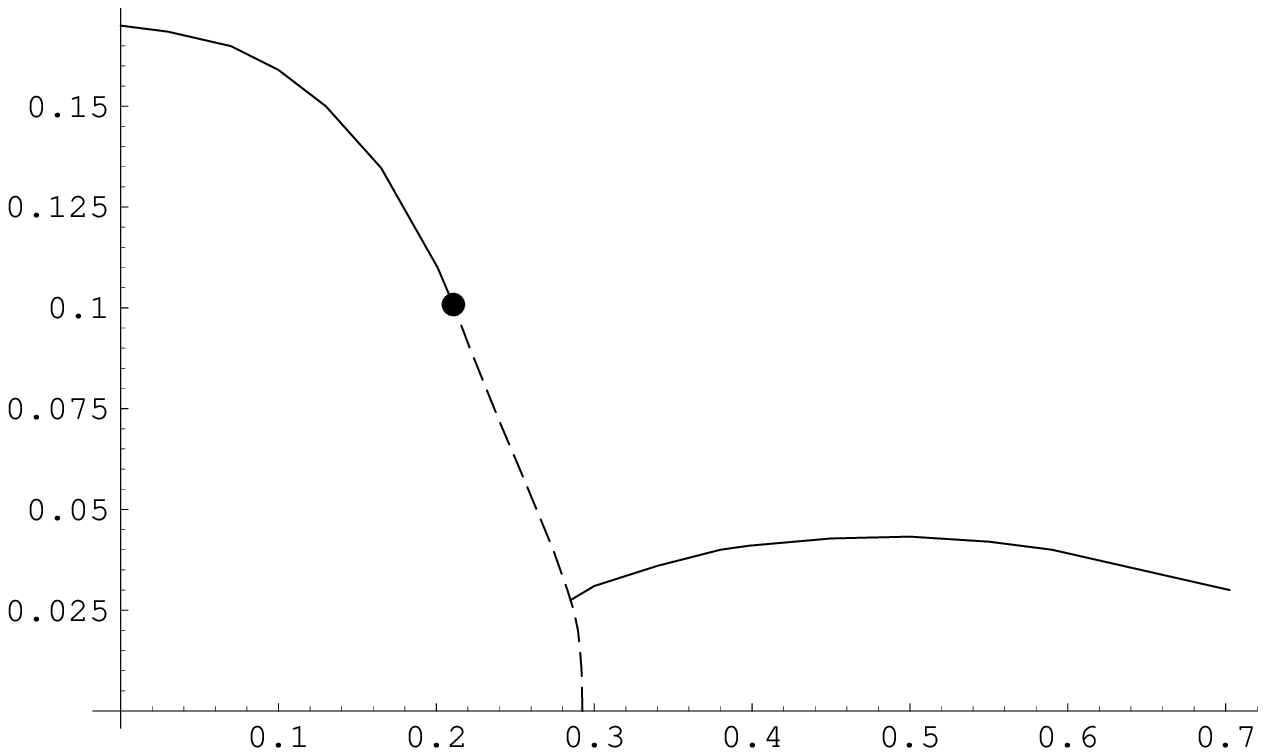}}
  \put(0.0,-0.6){
  \epsfysize=7.8cm
  \epsfbox[80 480 460 720]{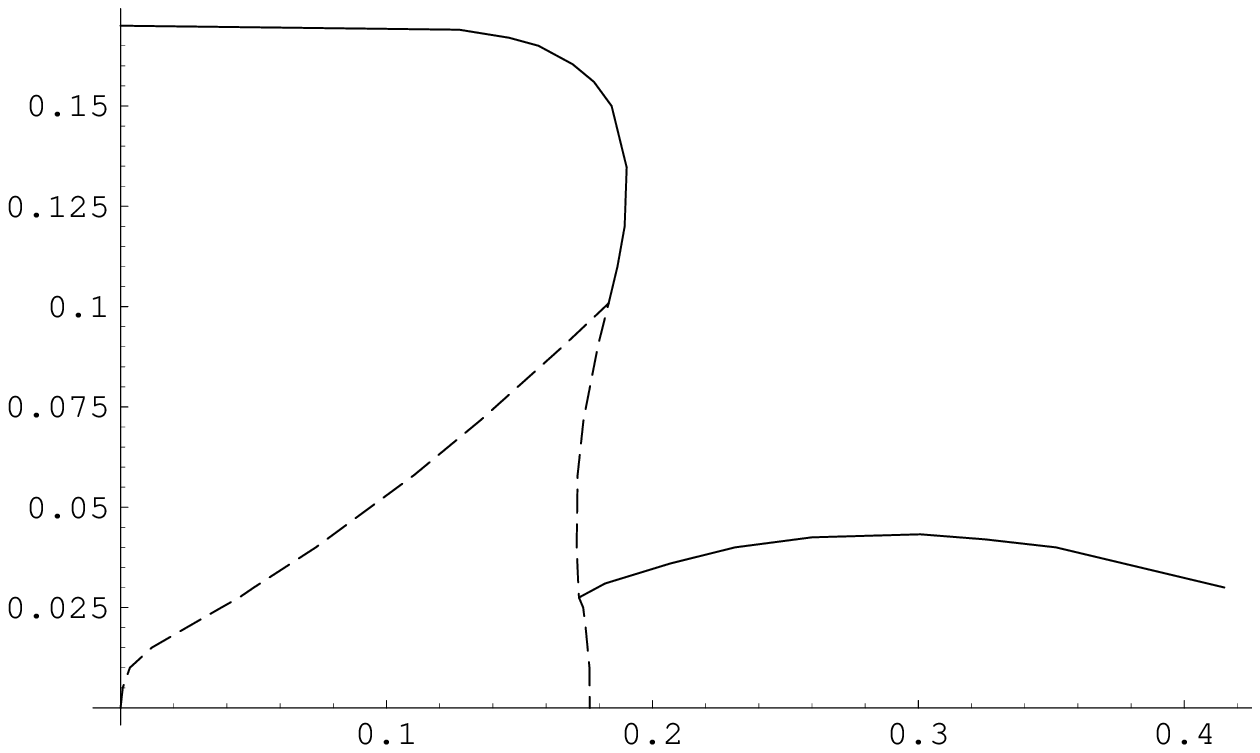}}
  \put(2.5,10.5){\bf $\Big{\langle}\bar{\psi}\psi \Big{\rangle}\not =0$}
  \put(7.8,9.2){\bf $\Big{\langle}\bar{\psi}\psi \Big{\rangle}=0$}
  \put(7,13.3){\bf $\Big{\langle}\bar{\psi}\psi \Big{\rangle}=
  \Big{\langle} \psi\psi \Big{\rangle}=0$}
  \put(2.5,9.8){\bf $\Big{\langle} \psi\psi \Big{\rangle}=0$}
  \put(7.8,8.5){\bf $\Big{\langle} \psi\psi \Big{\rangle}\not =0$}
  \put(5.1,11.9){tricritical point}
  \put(0.3,14.4){\bf\large $T$}
  \put(12.8,8){\bf\large $\mu$} 
  \put(2.5,4.8){\bf $\Big{\langle}\bar{\psi}\psi \Big{\rangle}\not =0$}
  \put(7.8,1.5){\bf $\Big{\langle}\bar{\psi}\psi \Big{\rangle}=0$}
  \put(7,5.3){\bf $\Big{\langle}\bar{\psi}\psi \Big{\rangle}=
  \Big{\langle} \psi\psi \Big{\rangle}=0$}
  \put(2.5,4.1){\bf $\Big{\langle} \psi\psi \Big{\rangle}=0$}
  \put(7.8,0.8){\bf $\Big{\langle} \psi\psi \Big{\rangle}\not =0$}
  \put(6.4,4.1){tricritical point}
  \put(3,1){mixed phase}
  \put(0.3,6.7){\bf\large $T$}
  \put(12.8,0.3){\bf\large $n^{1/3}$} 
  \end{picture}
  \end{center}
\vspace{-0.1in}
\caption{Phase diagram as a function of $T$ and $\mu$,
and as a function of $T$ and $n^{1/3}$ (all in GeV).  This section can be
viewed as one long caption for these figures.
The solid curves are second order phase transitions; the
dashed curves describe the first order transition.}
\vspace{-0.1in}
\end{figure}
In Figure 4, we present the phase
diagram of the model as a function of $\mu$ and $T$, and
as a function of $n$ and $T$.  
We devote this section to a discussion of the physics 
shown in these figures.
Let us first focus on
Figure 4a, and ignore the quark number
density $n$.  We find that at $\mu=0$, the chiral phase
transition as a function of increasing temperature is
second order.  In our present mean field treatment, it is of course
a mean field transition.  A second order chiral 
phase transition in two flavor
QCD at nonzero temperature 
has the appropriate symmetry to be
in the $O(4)$ universality class\cite{piswil,wil,RW,review},
and a treatment of the present model which went beyond mean
field theory, like that in Ref. \cite{BJW}, would find critical exponents
characteristic of this universality class. 
There is no reason to expect a small chemical potential to 
change this result, since this introduces no new massless
degrees of freedom in the effective three dimensional theory which
describes the long wavelength modes near $T=T_c$.
This argument has been made at greater length in Ref. \cite{hsu}.
However, it is possible that
as a parameter in the theory is changed, the quartic coupling
in the effective three dimensional theory can become
negative, making the transition first
order. Where the quartic coupling is zero, one has a tricritical
point.  (This is precisely what is expected to occur at $\mu=0$
at a particular value of the strange quark 
mass.\cite{wil,RW,review,GGP})  
Since we have seen that the zero temperature, nonzero $\mu$
transition is first order and we have found a second order 
transition at high temperature, we expect and find a tricritical
point where the second order 
transition becomes first order.
The tricritical point
occurs at $\mu=\mu_{\rm tc}=211$ MeV and $T=T_{\rm tc}=101$ MeV for the 
parameters we are using.

We have verified that 
at the tricritical point $\Omega\sim\phi^6$ as we expect,
based on the arguments above.\footnote{In fact,
$\Omega(\phi,\Delta=0;\mu_{\rm tc},T_{\rm tc})=0.033\phi^6 - 0.000392$, 
all in GeV.}
In the vicinity of
the tricritical point, $\Omega$ therefore has the form
\begin{equation}
\Omega(\phi,0;\mu,T)= \Omega(0,0;\mu_{\rm tc},T_{\rm tc})+
\frac{a(\mu,T)}{2} \phi^2 + \frac{b(\mu,T)}{4} \phi^4
+ \frac{c(\mu,T)}{6}\phi^6 - h\phi \ , 
\label{omegatc}
\end{equation}
where the
coefficient $h$ of the linear term is proportional to the current quark mass.
The coefficients $a$ and $b$ are both zero at the tricritical
point, and barring accidental cancellations both will be
linear in both $(T-T_{\rm tc})$ and $(\mu-\mu_{\rm tc})$.
The $\mu$ and $T$ dependence of $c$ is not important,
as $c$ does not vanish at the tricritical point, and it
is convenient to set $c=1$.
It is easy to verify that, near the
tricritical point, the line of second order transitions
is given by $a=0$, $b>0$ and the line of first order 
transitions is given by $a=3b^2/16$, $b<0$.
Minima of $\Omega$ are described by the scaling form
\begin{equation}
h=\phi_0^5\,\left( \frac{a}{\phi_0^4} + \frac{b}{\phi_0^2} + 1 \right)
\label{eqofstate} \, .
\end{equation}
{}From this, we read off the exponents $\delta=5$, $1/\beta=4$,
and $\phi_t/\beta=2$ where $\phi_t$ is called the crossover 
exponent because tricritical (as opposed to first or second
order) scaling is observed for $b<a^{\phi_t}$. For more
details see Refs. \cite{tricrit,review}. 
The relation between $a$ and $b$ and $(T-T_{\rm tc})$ and 
$(\mu-\mu_{\rm tc})$ depends on the slope of the line of
phase transitions which end at the tricritical point in Figure 4a 
and is not universal.
In general, our model should not be used quantitatively.
However, if we trust the qualitative feature that the
transition is second order at high temperatures and first order
at low temperatures, then QCD with two flavors will
have a tricritical point in the same universality 
class as that in our model. 
Since $d=3$ is the critical dimension for 
$\phi^6$ theory, we expect that 
a complete treatment 
would yield mean field exponents in agreement to
those in our model, although there would be
logarithmic corrections to the scaling law (\ref{eqofstate}) \cite{tricrit}.  
The reason we have described the physics of the
tricritical point carefully is that, because
we have a $\phi^6$ theory in three dimensions,
the universal critical exponents will be given
quantitatively by those we find in
mean field theory.

At low temperatures, 
one has color superconductivity  at chemical potentials 
immediately above the first order phase transition.
However, the superconducting order parameter $\Delta$
is reduced by nonzero temperature, as shown in
Figure 3. 
In Figure 4 we see that for the parameters
we are using, color superconductivity persists up
to temperatures as high as 45 MeV for the most favorable
chemical potential.  The energy gap in the color superconductor
phase (that is, the energy to create a quasiparticle--quasihole
pair) is $2\Delta_0 F(\mu)^2$.
BCS theory predicts that, at
weak coupling, the critical temperature $T_c^\Delta$
at which the superconductivity vanishes should be given by 
$0.57 F(\mu)^2\Delta_0(T=0)$. Our results are within
a few percent of this.  This suggests that $T_c^\Delta$
is sensitive to the shape of the form factor. For example,
we have verified that
if we use a form factor with a sharp cutoff we can
easily increase $T_c^\Delta$ by a factor of two, although
this tends to increase $T_c$ for the finite temperature
transition at zero chemical potential also. 
If we generalize beyond the instanton interaction,
and therefore do not require (\ref{g2g1}) but
simply impose $G_2<2G_1$ as required if color $SU(3)$
is to be a symmetry of the vacuum, then we find that 
values of $\Delta_0$ and $T_c^\Delta$ can arise which are
up to four times larger than those we show in our figures.

The transition at nonzero temperature at which color
superconductivity is lost is second order in our model.
Because of the shape of the $T_c^\Delta$ curve as 
a function of $\mu$, there are temperatures (like 
$T=30$ MeV in Figure 3) for which,
as $\mu$ is increased, one first completes the first
order chiral transition, then has a region with $\phi_0=\Delta_0=0$,
then has a second order transition to a color superconducting
phase, before finally having another second order transition
to a phase with no condensates.
If one neglects gluons, one could study this second order
transition by doing a renormalization group analysis 
to first determine whether there is a second order transition
as in our mean field analysis, and then to determine
the universal exponents due to the long
wavelength fluctuations of $\Delta$, associated with 
the breaking of the global color symmetry. 
We expect, however, that the second order nature of
the transition at $T=T_c^\Delta$ can change once gauge fields
are introduced.  
For example, the electroweak phase transition is
second order when only the Higgs field is included, but can
be either first order or a smooth crossover once gauge fields
are taken into consideration.  We leave a Ginzburg-Landau
treatment of the 
physics near $T_c^\Delta$ (including both gluons and the superconductor
order parameter $\Delta$) to future work.

By not showing $n$, Figure 4a obscures important
features seen in Figure 4b, including for example the fact 
(as discussed above) that 
at zero temperature, one of the phases in equilibrium at the
first order transition has zero density.  At any nonzero temperature 
which is low enough that the transition is first order,
the transition occurs between a gas of constituent quarks
with density $n_-$ and a gas of massless quarks with density
$n_+$.  In other words, in the mixed phase one has droplets
containing massless quarks as before, but they are now surrounded
by a dilute gas of constituent quarks, rather than by vacuum.
Once gluons are introduced, both the droplets of the
$n=n_+$ phase and the constituent quarks in the low
density phase are expected to form color singlets.
Thus, the description of
baryons as droplets is less complete at nonzero temperature
than at zero temperature.

It is clear that there is no model-independent
significance to the dashed lines in Figure 4b which mark
the boundaries of the mixed phase.  In nature, with
gluons and with repulsive short range
interactions between nucleons, the only curve which 
can be defined is the one which bounds the region 
in which chiral symmetry is broken.  Locating this
boundary requires further analysis of the first order
transition, but it seems likely\cite{ARW,ykis97,satz} that
it is associated with the percolation of the droplets
of the high density phase.  More precisely, once one no longer
has regions of the low density phase (with nonzero $\phi_0$)
which are of infinite extent, chiral symmetry will be restored.
Thus, the density at which chiral symmetry is restored 
must be somewhat to the left of $n_+$. Similarly,
color superconductivity sets in at densities above which
there are regions of the high density 
phase of infinite spatial extent. Chiral symmetry restoration
is associated with the ``unpercolation'' of the low density
phase, while the onset of color superconductivity is associated with the
percolation of the dense phase, and both occur at densities
somewhat below $n_+$.
None of this physics is 
visible in Figure 4a, since it all occurs within the mixed
phase region.  It is important to realize, therefore, that
if one is interested in the experimentally relevant question
of whether or not a first order transition occurs as $n$
is increased above nuclear matter density, this analysis
does not provide an answer.  The percolation transition could
be smooth or second order, even though it is occuring ``within''
the first order phase transition described by the model.
What is robust, and as we shall see 
in the next section may be quite significant,
is that regardless of how the physics as a function of $n$
works within the mixed phase region, there is a tricritical
point in the phase diagram.  

\section{Physics Away from the Chiral Limit}

\begin{figure}[t]
  \unitlength1.0cm
  \begin{center}
  \begin{picture}(13.0,5.0)
  \put(-2.0,0.2){
  \epsfysize=5.2cm
  \epsfbox[80 480 460 720]{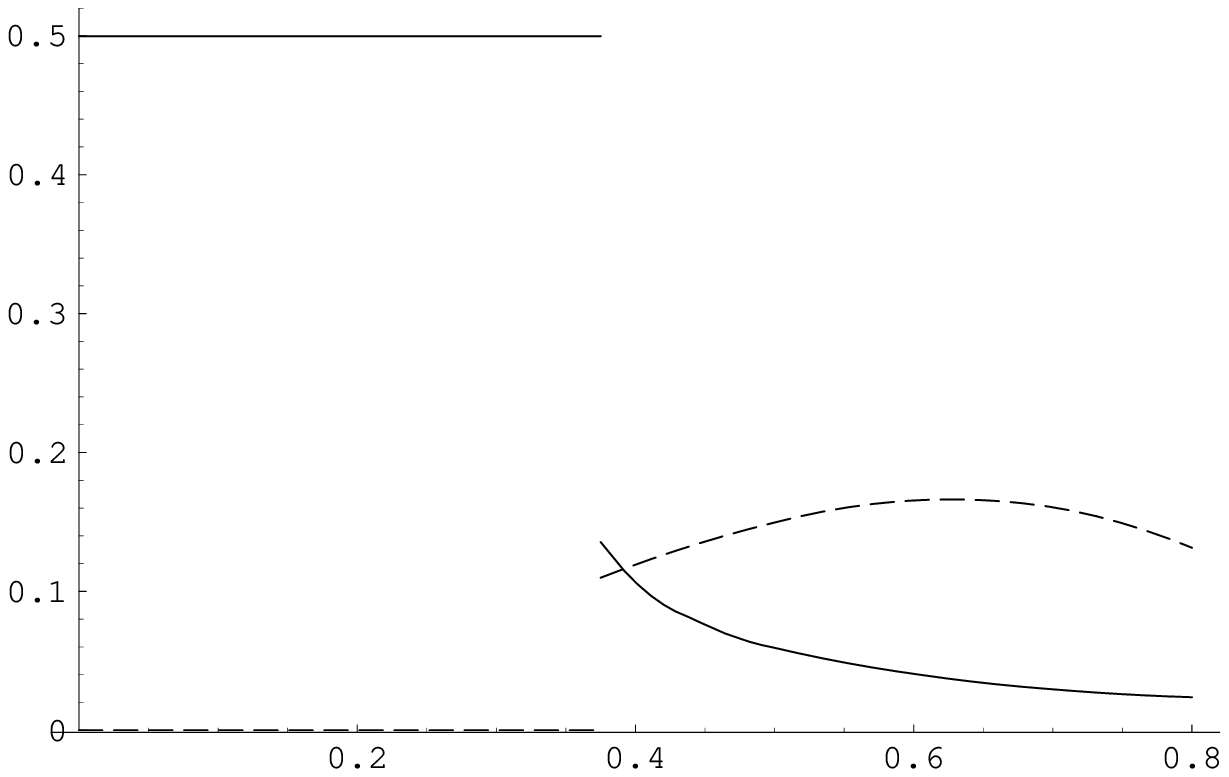}}
  \put(6.1,0.2){
  \epsfysize=5.2cm
  \epsfbox[80 480 460 720]{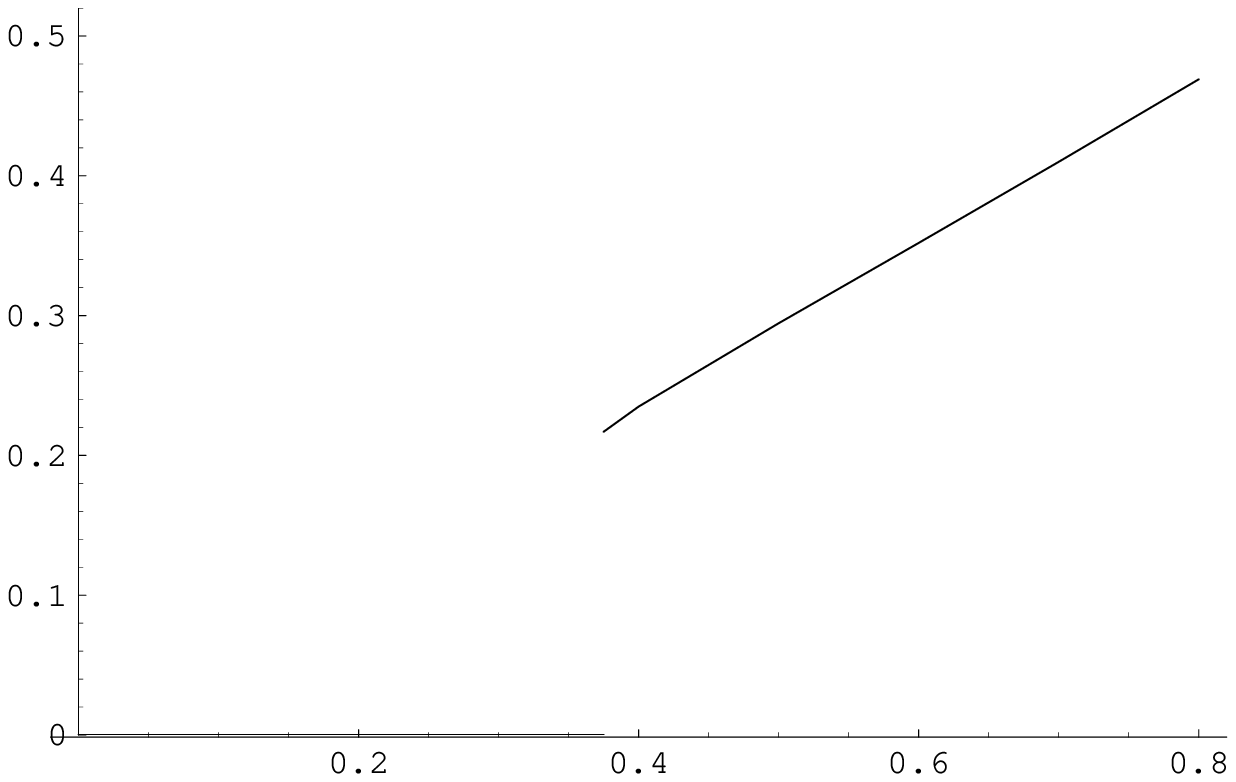}}
  \put(7.5,4.9){\bf $n^{1/3}$}
  \put(0,4.8){\bf $\phi_0$}
  \put(3.7,2.5){\bf $\Delta_0$}
  \put(2.5,0.2){\bf $\mu$}
  \put(10.6,0.2){\bf $\mu$}
  \end{picture}
  \end{center}
\caption{As for Figure 3
at zero temperature, but with a current quark mass $m=10$ MeV.
We observe coexisting $\phi_0 \sim \langle \bar \psi \psi \rangle$ 
and $\Delta_0 \sim \langle \psi \psi \rangle$ condensates in
the high density phase.}
\end{figure}
In the previous section, we have explored the phase diagram
at nonzero temperature and baryon density in the chiral limit.
We now turn on a small but nonzero quark mass $m$, as
in nature. 
We see from 
Figure 5 that in the low temperature phase, the quark mass
affects 
$\phi_0$, but leaves $\Delta_0=0$.  
We observe that $\phi_0^{\rm vac}$, $\mu_0$ and
the critical temperature for the chiral transition at 
zero density all increase with $m$.
One might reasonably choose to reduce $G_1$ as $m$ is
increased in order to keep $\phi_0^{\rm vac}$ fixed.  With $G_2/G_1$ fixed,
this would reduce $G_2$ and would therefore
reduce $\Delta_0$.
We do not do this because
our goal is to find any effects of the presence of the
chiral condensate on the superconductor condensate and
we do not wish these effects to be entangled with the 
effects of a reduction in $G_2$.
In the phase with density $n_0=(0.217\ \mbox{GeV})^3$ 
attained immediately upon
completion of the phase transition, both condensates
have comparable magnitudes with $\phi_0=0.136$ GeV
and $\Delta_0=0.110$ GeV.  We observe only a very small
dependence of $\Delta_0$ on $m$, as long as the constituent
quark mass $m+\phi_0$ is smaller than $\mu$.
(Compare Figure 5 to Figure 3, where $m=0$.) This insensitivity can be
understood by noting that although the Cooper pairs
have low momentum, they are formed from
quarks which
have momenta close to the Fermi surface.
Adding a quark mass $m\ll \mu$ 
does not significantly affect the density of states 
or the interactions of the quasiparticles with momenta
of order $\mu$. This is consistent with our finding
that the temperature
$T_c^\Delta$ at which the superconductivity
is lost is almost unchanged.

The phase transition at 
$T=T_c^\Delta$ 
remains second order in the presence of a nonzero quark mass.  
This result can be understood by
expanding the effective potential at small $\phi$ and $\Delta$. 
Whereas a quark mass adds a term to the effective potential
which is linear in $\phi$ at small $\phi$, it does not
introduce a term linear in $\Delta$.  It can 
modify the coefficient of the $\Delta^2$ term,  which
changes the critical temperature for the transition
but leaves it second order.

To conclude our discussion of color superconductivity, 
we remark that the phenomenon seems quite robust. It
is not significantly disturbed by the simultaneous 
presence of a chiral condensate as long as the constituent
quark mass is less than the chemical potential.  
Although the phenomenon is robust, our estimate
of the critical temperature at which it vanishes
is less so.  For the parameters we have used for
illustrative purposes, $T_c^\Delta$ is 30--45 MeV,
but changes in the form factor can double this and
phenomenologically acceptable changes in the coupling constant $G_2$
can multiply $T_c^\Delta$ by up to a factor of four.
A definitive answer of whether or not color superconductivity
can arise in heavy ion collisions, in which the necessary
densities are likely accompanied by temperatures of order
$100$ MeV, must await a less qualitative treatment.

Our model reveals the possibility of a tricritical point
in the phase diagram for two-flavor QCD in the chiral
limit.  Physics in the vicinity of the tricritical point
is described by a $\phi^6$ field theory, whose critical
exponents are given quantitatively by the mean field
analysis in this paper.  What, then, happens in this
region of temperature and chemical potential in the
presence of a small quark mass?
It is well known that a nonzero quark mass does have 
a qualitative effect on the second order transition 
which occurs at temperatures above $T_{\rm tc}$.  This $O(4)$ transition
becomes a smooth crossover, because terms linear
in $\phi$ now arise in $\Omega$.
A small quark mass cannot eliminate the first order
transition below $T_{\rm tc}$.  Therefore, whereas we previously had a line
of first order transitions and a line of second order
transitions meeting at a tricritical point, 
with $m\neq 0$ we now
have a line of first order transitions ending at an
ordinary critical point, as in the liquid-gas phase diagram.
At this critical point, one degree of freedom (that associated
with the magnitude of $\phi$) becomes massless, while
the pion degrees of freedom are massive
since chiral symmetry is explicitly broken.  Therefore,
this transition is in the same universality class as
the three-dimensional Ising model. 

{}From many studies of QCD at nonzero temperature, we are
familiar with the possibility of a second order transition,
with infinite correlation lengths, in an unphysical world
in which there are two massless quarks.  
It is exciting
to realize that if the finite density transition is first
order at zero temperature, as in the
models we have considered, then there is a tricritical
point in the chiral limit which becomes an Ising second
order phase transition in a world with chiral symmetry
explicitly broken.  In a sufficiently energetic heavy ion collision, one 
may create conditions in approximate local thermal
equilibrium in the phase in which spontaneous chiral symmetry breaking
is lost.
Depending on the initial density and temperature, when
this plasma expands and cools it will traverse the 
phase
transition at different points in the ($\mu,T)$ plane.
Our results suggest that
in
heavy ion collisions in which the chiral symmetry breaking
transition is traversed at baryon densities which
are not too high and not too low, 
a very long correlation length in the $\sigma$ channel
and critical slowing
down may be manifest even though the pion is massive.

\acknowledgments

We are especially grateful for discussions with M. Alford and F. Wilczek 
and thank R.~L.~Jaffe, D.-U. Jungnickel, J. W. Negele, D. T. Son,
M. Stephanov and C. Wetterich for very
useful conversations.

\end{document}